# Substrate-Induced Cooperative Effects in Water Adsorption from Density Functional Calculations

Pepa Cabrera-Sanfelix[1], M. V. Fernández-Serra[2], A. Arnau[1,3,4] and D. Sánchez-Portal[1,3]

[1] Donostia International Physics Center (DIPC), Paseo Manuel de Lardizabal 4, San Sebastian 20018, Spain

[2] Physics and Astronomy, State University of New York, Stony Brook, NY 11794-3800, USA

[3] Centro de Física de Materiales CSIC-UPV/EHU, Materials Physics Center MPC, Paseo Manuel de Lardizabal 5, San Sebastián 20018, Spain

[4] Departamento de Física de Materiales UPV/EHU, Facultad de Química, Apdo. 1072, San Sebastián 20080, Spain

Density Functional Theory calculations are used to investigate the role of substrate-induced cooperative effects on the adsorption of water on a partially oxidized transition metal surface, O(2x2)/Ru(0001). Focussing particularly on the dimer configuration, we analyze the different contributions to its binding energy. A significant reinforcement of the intermolecular hydrogen-bond (H-bond), also supported by the observed frequency shifts of the vibration modes, is attributed to the polarization of the donor molecule when bonded to the Ru atoms in the substrate. This result is further confirmed by our calculations for a water dimer interacting with a small Ru cluster, which clearly show that the observed effect does not depend critically on fine structural details and/or the presence of co-adsorbates. Interestingly, the cooperative reinforcement of the H-bond is suppressed when the acceptor molecule, instead of the donor, is bonded to the surface. This simple observation can be used to rationalize the relative stability of different condensed structures of water on metallic substrates.

## Introduction

The atomic level understanding of the adsorption of water on metallic substrates and, in particular, on Ru(0001), has received a lot of attention in recent years.[1-12] At large water coverage, the determination of the most stable structures of the water adlayer becomes a quite challenging task. This is mainly due to the interplay between two interactions of similar strengths: the intermolecular hydrogen-bond (H-bond) and the water-metal interaction.[3,10] In the case of Ru, an additional complication comes from the competing stability of the partially-dissociated and the intact-molecule adsorption configurations of water.[5,10,12]

Using scanning tunnelling microscopy (STM) supplemented by density functional (DFT) calculations, Michaelides and Morgenstern[13], were recently able to resolve the structures of small water clusters adsorbed on Cu(111) and Ag(111). These results help to characterize the ability of water molecules to simultaneously bond to a metallic substrate and to form H-bonds between them and, therefore, to rationalize the observed structures for extended H-bonded water networks on metallic substrates. From a structural point of view, there are at least two types of water molecules: one type forms a direct bond with the substrate, whereas the other is essentially ice-like H-bonded and has little interaction with the metal substrate.[14]

From an energetic point of view, the increase of the water coverage on metallic substrate, up to the complete monolayer, usually enhances the adsorption energy per molecule. This is easily understood since the coordination (and the average number of H-bonds per molecule) increases as extended water networks growth on the substrate.[7, 10] This also occurs on ionic[15] and graphitic substrates.[16] However, the formation of extended water networks is typically accompanied by other phenomena that also cause an appreciable increase of the binding energy per water molecule. These are the so-called *cooperative effects* in the water-water interaction: the strength of the H-bond between two water molecules is largely increased by the fact that those molecules have additional H-bonds with other neighbouring molecules. In general, other long range electrostatic interactions can also favour lager polarization of the water molecules and induce a strengthening of the H-bonds.

The cooperative or non-pairwise character of the interactions is, for example, a fundamental property of liquid water, where H-bonds are up to 250% stronger than for the isolated H-bond of the dimer.[17] Cooperative effects are also quite strong in small water clusters.[11, 18-24] The key ingredient to understand these cooperative effects is the polarization induced in each water molecule by the presence of their neighbours. This polarization has its major effect on the lone pairs of the oxygen atoms, which are the electrons involved in the formation of hydrogen bonds. The donor O-H covalent bond also suffers a significant polarization with a net shift of the electron density toward the oxygen of the donor molecule. In the nomenclature used throughout the paper, donor molecule refers to the water molecule whose Hydrogen atom is pointing to and interacts with the electronegative Oxygen atom of the acceptor water molecule. These effects lead to the increase of the molecular dipole[26] and, correspondingly, to the strengthening of the new H-bonds formed with additional water molecules. The quantum character of the protons also contribute to enhance these effects, leading to the elongation of the O-H covalent bond in condensed phases and a further increase of the dipole moment.[27]

On forming the H-bond, the donor hydrogen atom moves away from its oxygen and the acceptor lone-pair stretches toward the donor hydrogen. Thus, both oxygen atoms are pulled together while the covalent O-H bond is being stretched and weakened.[28] The main origin of this weakening is the Pauli repulsion between the lone pair of the acceptor molecule and the electrons localized on the O-H bond of the donor molecule.[29] Additionally, the formation of the H-bond gives rise to a small (few milielectrons)[1, 12, 30] charge transfer from the lone pair of the acceptor molecule to the donor molecule. This transfer also contributes to the weakening of the O-H covalent bond of the donor molecule.[31] These effects are the reason, for example, for the red shift of the O-H stretching frequency in ice versus liquid water. This shift also correlates with a blue shift of the H-bond stretching band in ice as compared to water. So there is a correlation between the strength of these two bonds: the stronger is the H-bond, the weaker the covalent O-H bond and the shorter the distance O⋯O between the oxygen atoms of both molecules. Therefore, the weakening and elongation of the O-H covalent bond becomes a good indicator of the stability of the H-bond.[32] The main observables that can be correlated to the strength of H-bonds are the intermolecular distances and the frequency shifts of the stretching modes of those covalent bonds containing the donor hydrogen atoms.

Much work to date has been devoted to the study of cooperative effects in water networks. However, we can also expect the appearance of substantial cooperative effects in other situations. The necessary condition is that a strong polarization is induced in those water molecules participating in the H-bond and, in particular, in the donor molecule. The adsorption of water on some substrates can provide a mechanism to generate such additional polarization. Indeed, as we will see below, the adsorption of the donor water molecule to some substrates, characterized by a strong oxygen-metal interaction, can give rise to a significant strengthening of the H-bond, comparable to that observed in water and ice.

In the present work, we investigate in detail the inter-molecular H-bond in a water dimer interacting with different substrates. Our initial motivation comes from the observation of anomalously large adsorption energies per molecule for a water dimer adsorbed on O(2x2)/Ru(0001).[33] This structure was initially proposed to explain some of the STM images obtained for water deposited on this substrate at coverages larger than 0.25 ML. Here we perform a quantitative analysis of the different contributions to this large binding energy and conclude that it is mainly due to the strengthened H-bond within the water dimer. The shifts of the calculated vibrations and, in particular, that of O-H stretching mode are used to characterize this increase in stability. Afterwards, we use a simple model to explore the effect of adsorption on a general Ru substrate: a water dimer interacting with a small metal cluster. This simple model confirms the validity of our initial conclusions and demonstrates that H-bond stabilization stems primarily from the interaction of the donor molecule with the metallic substrate, and does not depend critically on fine structural details or the presence of co-adsorbates. Indeed, our work seems to indicate that the appearance of *strong cooperative effects* in the water-water interaction induced by the adsorption to a substrate is quite general. This seems to be confirmed, for example, by the high binding energy per molecule obtained for a water dimer adsorbed on an ionic substrate like NaCl.[15, 34]

**Theoretical Method**
Our DFT calculations were performed using the Vienna package (VASP),[35-37] within the Perdew-Wang 1991 (PW91) version of the general gradient approximation (GGA).[38] The projector augmented wave (PAW)[39, 40] method was used to describe the interaction of valence electrons with the Ru, O and H cores. To describe the Ru(0001) and O/Ru(0001) surfaces we used a symmetric slab containing seven Ru layers plus a similar amount of vacuum between periodic replicas of the slab. Adsorbates were always placed on both sides of the slab to keep the mirror symmetry and the zero total dipole moment along the normal to the surface, consistent with the periodic boundary conditions. These computational details are similar to those used in our previous work on similar systems.[12, 33] A plane-wave cutoff of 400 eV and a 6x6x1 k-point sampling was used for our smallest cell, corresponding to a 2x2 unit cell in the lateral directions. For the larger 4x4 unit cell, used to represent lower water coverage, the k-point sampling was reduced to 3x3x1. All geometries were optimised by allowing relaxation of all degrees of freedom of the two outermost Ru layers and the O and H atoms until residual forces were smaller than 0.03 eV/Å. This procedure has been proved to be accurate enough.[33] The adsorption energies of the water molecules are calculated as described in previous work.[33] Although it is well known that DFT does not account for dispersion interactions, H-bonds are reasonably well described using the PW91 DFT-GGA functional for exchange and correlation.[41] For presentation purposes, we have replaced the discrete Delta functions by Gaussian functions with a width of 1.5cm$^{-1}$ in the plots of vibration density of states (VDOS).

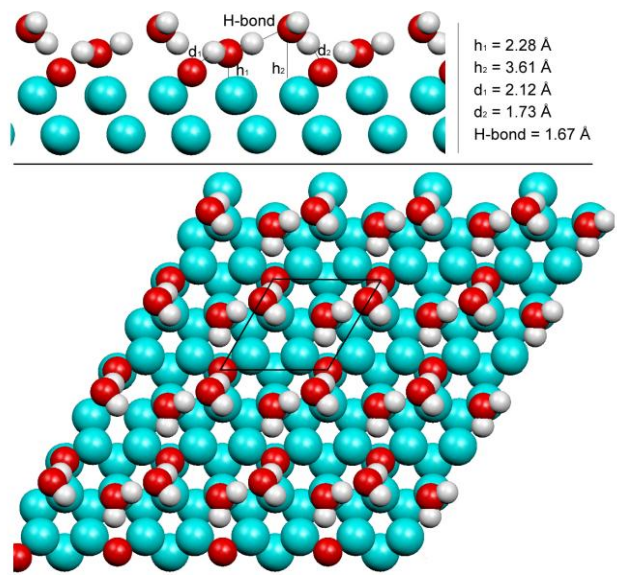

**Figure 1.** Calculated water configuration at 0.5 ML coverage on the O(2x2)/Ru(0001) surface. The structure is formed by dimers in which one of the molecules has its oxygen 2.26 Å above a Ru top site and it is hydrogen bonded to one of the surface oxygen atoms (2.12 Å bond length) and to the adjacent water molecule (very short H-bond of 1.67 Å). The second molecule adsorbs 3.61 Å above the Ru topmost layer, and forms a H-bond to the substrate oxygen atom right below (1.73 Å bond length).

**Results and Discussion**
*Water dimer on O(2x2)/Ru(0001)*
The saturated first hydration layer on clean Ru(0001) corresponds to a water coverage ($\theta_w$) of 2/3 ML and follows a commensurate structure,[42, 43] whereas 1/4 ML of water is sufficient to saturate the O(2x2)/Ru(0001) surface, following a well ordered p-(2x2) symmetry.[33] In both cases, the most favourable adsorption site for a water monomer is on Ru atop sites. In the case of the O(2x2)/Ru(0001) surface the hydrogen atoms of the molecule point toward the neighbouring pre-adsorbed O atoms on the surface to form two long H-bonds. Therefore, at 1/4 ML coverage all the preferred adsorption sites in O(2x2)/Ru(0001) are occupied.

However, in a recent work[33] we found that a very stable structure can be obtained for the O(2x2)/Ru(0001) surface at a larger coverage of 1/2 ML. In this case, only half of the water molecules are directly bonded to the metal atoms of the surface, while the other half is attached uniquely through H-bonds to the adsorbed water molecules and the pre-adsorbed O atoms on the surface. The resulting structure is based on the water dimer as a building-block (see Figure 1). Surprisingly, this geometry is degenerate with the saturated water overlayer at 1/4 ML in which all the molecules occupy preferred adsorption sites and are well attached to the metal. Furthermore, the isolated water dimer is also very stable on this surface, i.e., at much lower water coverage the dimer continues to be a very favourable adsorption configuration.

**Table I.** Structure and energetics of the free-standing and adsorbed water dimer. Distances in the schemes are given in Å. The binding energies associated with the different hydrogen bond are listed, for the dimer on O(2x2)/Ru(0001) we also show the adsorption energy per water molecule.

| System | Scheme | H-bond (meV) | $E_{ads}/H_2O$ (meV) |
|---|---|---|---|
| (a) Relaxed water dimer in vacuum | 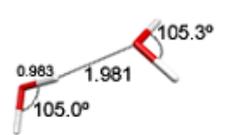 | 237 | --- |
| (b) Ads. water dimer on O(2x2)/Ru(0001) | 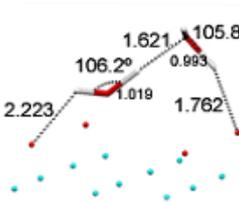 | 724 | 600 |

Figure 1 shows these stable water bilayer on O(2x2)/Ru(0001). The lower laying molecule is well attached to an available Ru atop site in the cell, its molecular plane slightly tilted and its orientation such that it optimizes a long H-bond with the surface oxygen (2.12 Å). The second molecule is located 3.61 Å above the surface and forms a H-bond (1.73 Å) with the surface oxygen atom underneath. The molecules stick together through a relatively short H-bond (1.67 Å), in which the molecule bonded to the metal acts as the hydrogen donor. The adsorption energy for this relaxed bilayer configuration is 600 meV per water molecule, comparable to the 590 meV found for the saturated layer at 1/4 ML.[33] At lower coverage the geometry of the water dimer maintains the main characteristics described above. This is the case of the structure described for $\theta_w \sim 1/8$ML in Ref.[32]. This diluted dimer is also energetically degenerate with structures formed by collections of optimally adsorbed monomers at the same coverage. The results for the energetics of the water dimer on O(2x2)/Ru(0001) are quite puzzling. On the one hand, according to our analysis, the adsorption energy of the water monomer on O(2x2)/Ru(0001) can be approximately divided in two main contributions: *(i)* the interaction with the metal substrate ($E_{metal}$) that contributes with ~340 meV and, *(ii)* the two long H-bonds ($E_{l-Hb}$) with the surface oxygens, each of them contributing with ~120 meV.[33] On the other hand, at the adsorbed dimer configuration only one molecule interacts directly with the metal, while two H-bonds are formed with the oxygen atoms on the substrate and one H-bond between the two molecules. The calculated H-bond for the free-standing dimer with our method is ~240 meV (see Table I). Thus, a rough estimation of the binding energy per water molecule for the adsorbed dimer structure gives $E_{ads} \sim (E_{metal} + 2xE_{l-Hb}) +240)/2 = 410$ meV/$H_2O$, which is almost 200 meV/$H_2O$ smaller than the actual calculated value. Even if the H-bonds formed with the substrate are in average more stable for the dimer than for the monomer (since they are significantly shorter in the former case), the estimated $E_{ads}$ per molecule would be at least ~100 meV too low. In the following, we study in detail the reasons for this high stability of the adsorbed dimer and conclude that the key ingredient is the strengthening of the inter-molecular H-bond due to the chemical interaction of the donor water molecule with the Ru metal atoms in the substrate.

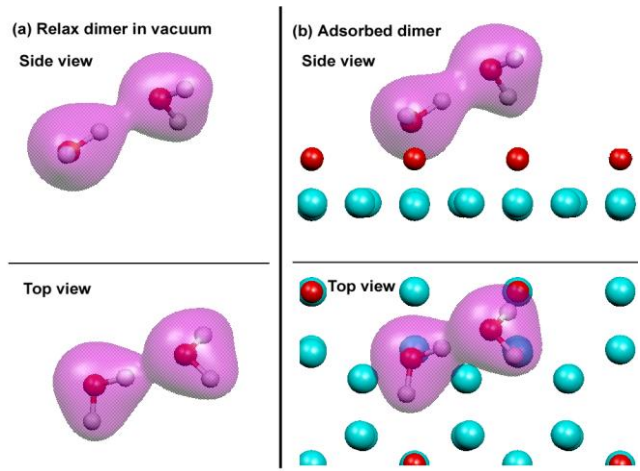

**Figure 2.** Isosurface of the calculated charge density for a water dimer:(a) free-standing water dimer ($\rho_{dimer}$); (b) adsorbed water dimer on the O(2x2)/Ru(0001) surface. Notice that in this second case we have subtracted the charge density corresponding to the surface, i.e., $\rho_{dimer} = \rho_{dimer/surface} - \rho_{surface}$. In both cases, the value of the density used to plot the surfaces is 0.02 a.u.

As a first step, we compare the geometry and energetics of the adsorbed and free-standing water dimer, both computed using a supercell of the same size. The results are presented in Table I. The calculated length of the H-bond for the free-standing dimer (1.89Å) is comparable, although slightly shorter, than that obtained with other theoretical methods like Hartree-Fock (HF/6-311++G**, 1.95 Å) or MP2 (MP2/6-

311++G**, 2.06 Å).[32, 44]. The H-bond is ~0.25Å shorter for the adsorbed dimer. This is an indication of the reinforcement of the intermolecular interaction. In water clusters, the stabilization of the H-bond network has been recognized to increase with the number of molecules forming the cluster.[19, 45] As mentioned in the introduction, the importance of such cooperative effects has been also recognized in liquid water.[17] In our case, the strengthening of the intermolecular H-bond should come from the interaction with the substrate.

In order to explore this effect, Figure 2 presents the charge density surface ($\rho_{dimer}$) for both, the free-standing dimer and the adsorbed dimer on the O(2x2)/Ru(0001) surface. Notice that in the latter case [Fig.2(b)] the charge density correspondent to the O(2x2)/Ru(0001) surface has been subtracted (i.e., $\rho_{dimer} = \rho_{dimer/surface} - \rho_{surface}$). The shapes of the density iso-surfaces are very similar in both cases. However, the wider section of the iso-surface in the intermolecular region for the adsorbed dimer is a clear signature of a larger charge accumulation in that region. This indicates that the interaction with surface indeed influences the inter-molecular H-bond.

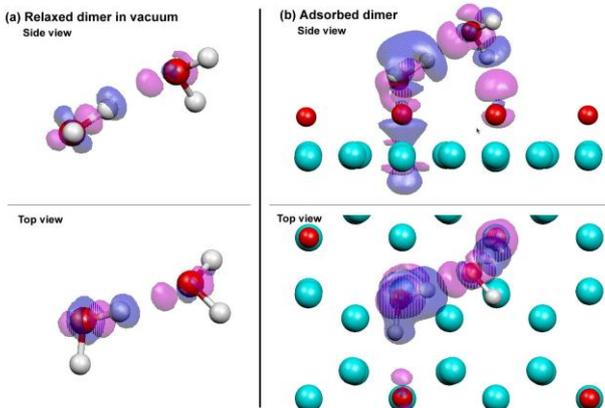

**Figure 3.** Calculated induced charge density for a water dimer: (a) free-standing dimer ($\rho_{ind} = \rho_{dimer} - \rho_{monomers}$); (b) adsorbed water dimer on the O(2x2)/Ru(0001) surface ($\rho_{ind} = \rho_{dimer/surface} - \rho_{surface} - \rho_{monomers}$). Light (pink) surfaces correspond to electron accumulations, while dark (blue) correspond to electron depletions. Surfaces are plotted for the isovalues ±0.003a.u..

These effects become more evident when plotting the induced charge density, as shown in Figure 3. The plot for the free-standing dimer clearly indicates that the formation of the H-bond is accompanied by the polarization of both molecules: electron depletion appears around the H atom of the donor molecule, whereas charge accumulation, pointing toward the neighbouring molecule, is visible on the O atom of the acceptor molecule. The adsorption of the dimer to the substrate creates a slightly more complex pattern with several contributions [Figure 3 (b)]. In addition to the inter-molecular H-bond, the additional bonds formed with the substrate also act as additional sources of polarization. In particular, the interaction of the donor molecule with the Ru atom underneath gives rise to an appreciable charge rearrangement characterized by a depletion of electron charge around the Ru atom and an electron accumulation on the oxygen of the molecule. Therefore, the polarization induced by the interaction of the donor molecule with the metal tends to enhance, rather than to cancel, the polarization pattern induced by the inter-molecular H-bond. It is interesting to notice that similar charge density rearrangements have been recently presented by Michaelides and Morgenstern for the adsorption of water hexamers on Cu(111).[13] In these buckled clusters only half of the molecules are well attached to the Cu atoms and, therefore, are reminiscent of the dimer structures considered here. According to Michaelides and Morgenstern, the substrate-induced polarization of these water molecules in the buckled-hexamer increases the adsorption energy per water molecule by more than 100 meV with respect to that of a forced planar-hexamer. Therefore, the results contained in Figure 3 are indicative of an important effect of the substrate in the polarization of the molecules and H-bond enhancement. We have, however, performed a more quantitative estimation of the influence of the substrate on the intermolecular H-bond, which is presented in what follows.

Table II presents the different contributions to the energetics of the water dimer adsorbed on the O(2x2)/Ru(0001) substrate. The intermolecular interaction in the isolated dimer keeping the geometry of the adsorbed configuration (i) is ~131 meV. The dimer-substrate interaction, (ii) ~1.15 eV, is estimated by subtracting the energies of the isolated dimer and of the O(2x2)/Ru(0001) surface, again both keeping the geometry of the combined relaxed system, from the total energy of the optimized adsorbed water dimer on O(2x2)/Ru(0001). Two additional single point total-energy calculations have been performed taking out each water molecule from the optimized adsorbed configuration (schemes (iii) and (iv) in Table II), while keeping the geometry of the rest of the system. The corresponding energy values represent the interaction of each of those water molecules with both, the substrate and the remaining molecule. Then, the strength of the intermolecular H-bond can be determined by adding both energies (iii) and (iv) and subtracting the dimer-substrate interaction (ii). With this procedure, the intermolecular H-bond has been estimated to contribute ~724 meV to the stabilization of the system. This value is much higher than the ~131 meV for the H-bond of the fixed-geometry dimer in vacuum or the 237 meV binding energy of the optimized free-standing dimer. Thus, the substrate induced polarization of the donor molecule makes the intermolecular H-bond of the adsorbed water dimer at least three times stronger than for the free-standing dimer in vacuum.

**Table II.** Analysis of the different contributions to the binding energy of the adsorbed water dimer on O(2x2)/Ru(0001).

| System | Scheem | Calculation | Energy (eV) |
|---|---|---|---|
| (a) Free-standing dimer in vacuum (fixed geometry) | | (i) Intermolecular interaction [(a) – two H₂O in vacuum] | 0.13 |
| (b) Adsorbed dimer on O(2x2)/Ru(0001) | | (ii) Dimer – substrate interaction [(b) – (a) – Substrate] | 1.15 |
| | | (iii) Interaction on molecule (1) with (2) and substrate [(b) – (2) – (1) + Substrate] | 1.19 |
| | | (iv) Interaction on molecule (2) with (1) and substrate [(b) – (2) – (1) + Substrate] | 0.69 |
| | | (v) Intermolecular H-bond [(iii) + (iv) – (ii)] | 0.72 |
| | | (vi) Substrate contribution to the intermolecular H-bond [(v) – (i)] | 0.60 |

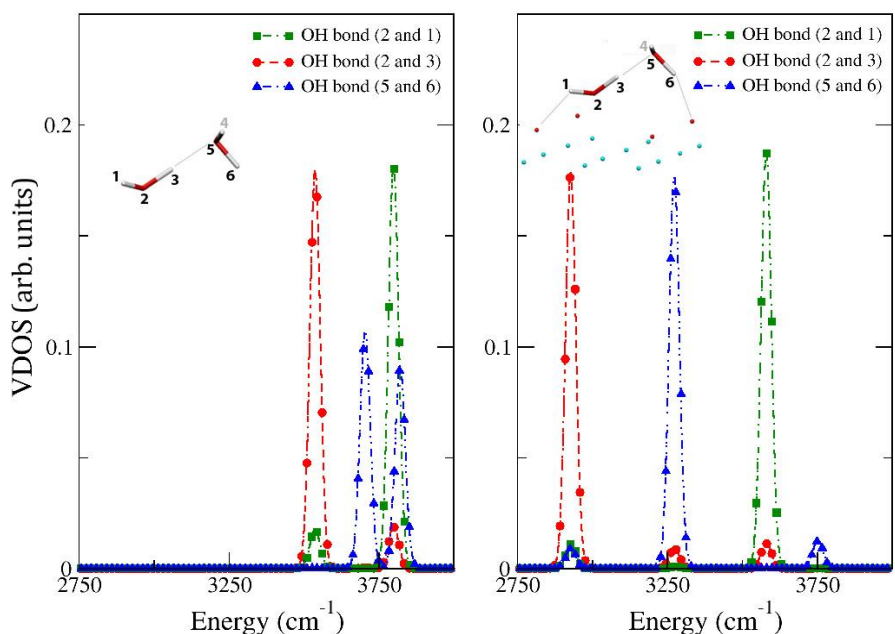

**Figure 4.** Vibrational density of states (VDOS) of the relaxed free-standing water dimer (left panel) and of the adsorbed dimer on the O(2x2)/Ru(0001) surface (right panel). Only the contributions to the VDOS from those H atoms involved in the H-bonds of the adsorbed dimer have been included in both panels.

The polarization of the water molecule, that strengthens the intermolecular H-bond, is also expected to cause a simultaneous weakening of the covalent O-H bonds within the molecule. Figure 4 shows the vibration density of states (VDOS) of both, the relaxed free-standing and the adsorbed water dimer on the O(2x2)/Ru(0001) surface. The calculated energy range of the stretching of O-H bonds for the free-standing dimer (3500 - 3800 $cm^{-1}$) is in good agreement with that found in the experiments.[21, 46] As expected, the stretching mode of the donating O-H covalent bond [between atoms 2 and 3 in Figure 4] has the lowest frequency, reflecting the weakening of this O-H bond. For the adsorbed dimer the stretching modes cover a larger frequency range (2950 - 3800 $cm^{-1}$). In particular, as a clear signature of the strong intermolecular H-bond, the frequency of the stretching mode of the donating O-H bond in this case shifts downward more than 700 $cm^{-1}$ respect to the stretching of those O-H bonds not involved in H-bonds. This finding is in agreement with the experimental results by Morgenstern and co-workers,[11] that observed using inelastic electron tunnelling spectroscopy a large red shift of ~750 $cm^{-1}$, compared to gas phase, of the OH stretching for extended structures of ice adsorbed on Ag(111). As pointed out by these authors this, red shift is too large to be only attributed to the increase of the coordination number of the water molecules in the adsorbed water layer. However, in the light of our calculations, we can interpret this measurement as an evidence of the important role of water-metal bond on the reinforcement of the intermolecular H-bond. The stretching modes of O-H bonds involved in the formation of other H-bonds with surface oxygens are also shifted down. For example, the stretching of the O-H bond between atoms 5 and 6 [see Figure 4 (b)] decreases by ~450 $cm^{-1}$ with respect to the free-standing case. This shift is even larger than the associated with the formation of the intermolecular H-bond in the free-standing dimer and is a signature of an important strengthening of the H-bond with the oxygen atoms on the substrate. This shows that, in addition to the binding with the metal, the fact that the molecule forms more than one H-bond also tends to increase the stability of the whole H-bonded network as already observed for free-standing water clusters.[21, 24]

An interesting point to consider is that the weakening of the covalent O-H bonds for the adsorbed water dimer and other buckled structures on the oxidized Ru substrate, associated with the formation of strong H-bonds, may have consequences for the chemistry of water. For example, a reduction of the energetic barriers for partial dissociation can be expected in some of these structures. The experimental observation that the presence of pre-adsorbed oxygen, at low coverage, on Ru(0001) promotes

the dissociation of water can support this interpretation.[2, 6]

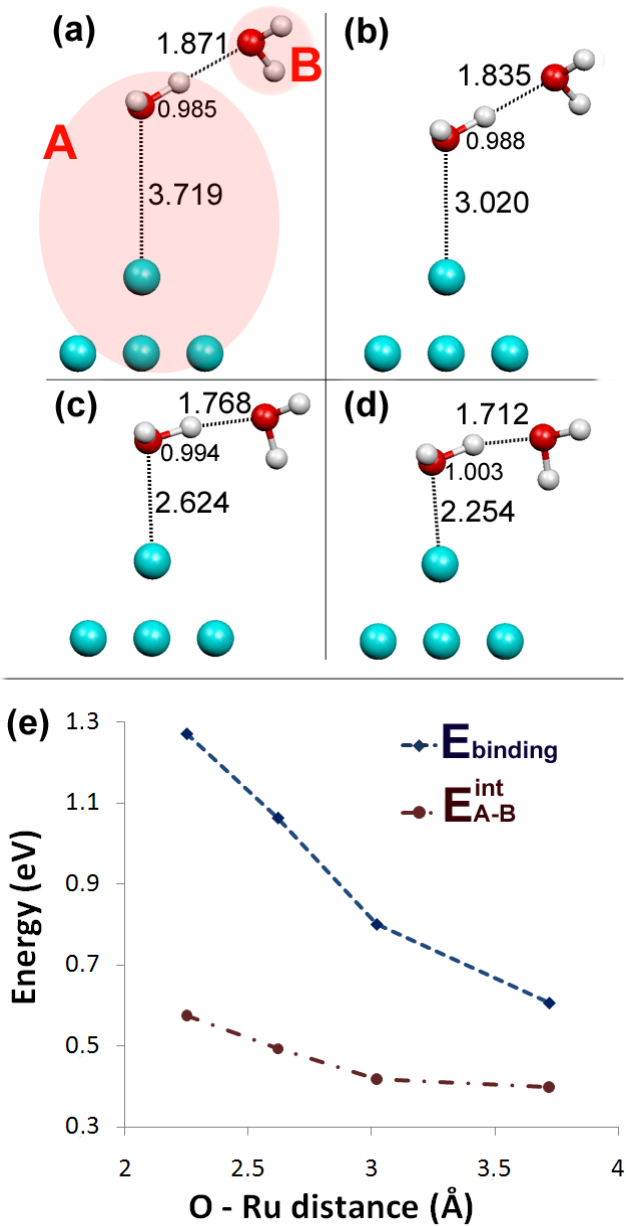

**Figure 5.** (a-d) Relaxed structures of a water dimer interacting with a small Ru cluster. The distance between the cluster apex and the oxygen atom of the donor molecule is kept constant during the relaxation. Case (d) corresponds to the optimum Ru-O separation. All distances are given in Å. Panel (e) shows the energetics of the system.

$E_{A-B}^{int}$ is the interaction energy of the fragments A (cluster+donor molecule) and B (acceptor molecule) defined in panel (a). The changes of $E_{A-B}^{int}$ ($\Delta E_{A-B}^{int}$) give a direct estimation of the H-bond strength variation with the Ru-O distance. $E_{binding}$ represents the total binding energy of the system respect to their components, i.e., the isolated cluster and two isolated water molecules.

*Role of the water-metal interaction*
Finally, we perform a meticulous analysis of the water-metal interaction on the observed reinforcement of the H-bond. In order to do this, we investigate the effect of approaching a small metal cluster, a Ru tetrahedron, to a water dimer in vacuum. At each step the metal atoms and the oxygen atom of the donor molecule remain fixed, while the rest of the atoms are allowed to relax. The distance ($d_{Ru-O}$) between the apex of the Ru cluster and the oxygen of the donor molecule are varied from 3.72 Å to 2.25 Å, the last value corresponding to the optimum Ru-O distance. The resulting geometries can be found in Figure 5 (a-d). The H-bond between the water molecules becomes shorter (up to a ~9.5% shorter) as the donor water molecule approaches the Ru cluster, being a clear signature of the H-bond reinforcement by the metallic cluster. This is confirmed by our energetic calculations illustrated in Figure 5 (e). The interaction energy for each configuration $E_{A-B}^{int}$ is calculated by comparing the energies from single point calculations of fragments A and B defined in Figure 5 (a) with that of the combined system. Fragment A comprises the cluster plus the donor water molecule, while fragment B is the acceptor water molecule. The changes on $E_{A-B}^{int}$ as the dimer approaches the Ru cluster can be used to estimate the change of the H-bond strength, i.e. $\Delta E_{A-B}^{int}$. $E_{A-B}^{int}$ increases up to ~180 meV as a result of the interaction of the donor molecule with the metal cluster. Comparing this variation of $E_{A-B}^{int}$ with the H-bond in a free-standing dimer (see Table I), we conclude that solely the interaction of the donor water molecule with the metal can already account for an increase of at least a 75% in the strength of the intermolecular H-bond.

Figure 6 (a-d) shows the induced charge density surface as we approach the water dimer to the Ru tetrahedron. The strong polarization of the system as $d_{Ru-O}$ decreases gives a good insight of the reinforcement of the dimer hydrogen bond. As the dimer comes closer to the Ru cluster the metal strongly polarizes the donor molecule. A growing electron depletion appears in the Ru atoms closer to the donor molecule, similar to what we already observed for the dimer on O(2x2)/Ru(0001) substrate [see Fig.3(b)]. Correspondingly, we observe the increasing polarization, and consequent weakening, of the donating O-H covalent bond. In turn, the polarization of the donor molecule also induces an additional polarization in the acceptor, as can be seen in the increasing electron accumulation surrounding the oxygen atom of that molecule. This combined effect reinforces the H-bond as the metal is approached.

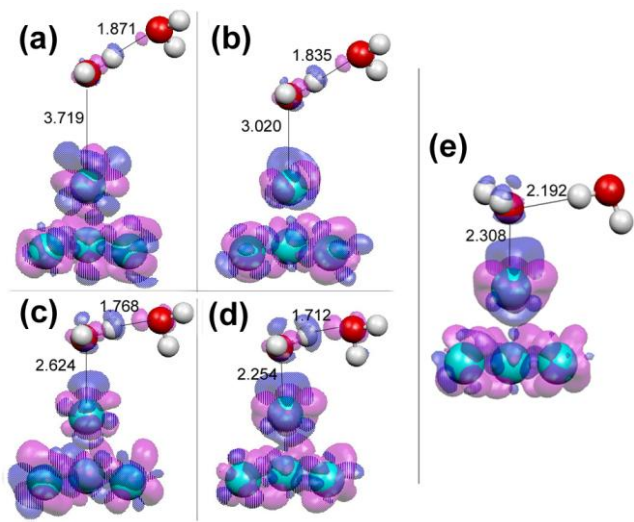

**Figure 6**. Panels (a) to (d): calculated induced charge density for the systems presented in Figure 5. The induced charge is calculated as $\rho_{induced} = \rho_{dimer/cluster} - \rho_{donor} - \rho_{acceptor} - \rho_{cluster}$, where $\rho_{dimer/cluster}$ is the electron density of the combined system, $\rho_{donor}$ and $\rho_{acceptor}$ are the densities of the isolated water molecules, and $\rho_{cluster}$ that of the isolated Ru cluster. Light (pink) surface correspond to the isovalue +0.004 a.u. (electron accumulation) and the dark (blue) surface correspond to isovalue -0.004 a.u.(electron depletion). Panel (e) shows the relaxed configuration when the dimer is interacting with the cluster through the acceptor molecule, along with the corresponding induced charge $\rho_{induced}$. All distances are given in Å.

We have checked that the large stabilization of the dimer is indeed related to the polarization of the donor molecule and the corresponding H-bond strengthening. If instead of the donor we approach the acceptor molecule to the metal cluster, the effect is the opposite and the dimer is destabilized (see Figure 6 (e)). For the optimum distance between the acceptor molecule and the cluster apex, the H-bond elongates to 2.192 Å (~17%) and the total binding energy of the system is reduced by ~350 meV compared with the optimum binding between the Ru cluster to the donor molecule. In this case, although the interaction with the metal also polarizes the acceptor molecule, the presence of the neighbouring metal efficiently screens the charge of the oxygen atom in the acceptor molecule and, therefore, the H-bond is strongly weakened. Thus, these calculations clarify the role of the substrate on the adsorption of water: strong substrate induced cooperative effects in the H-bond network of water are directly associated with the preferential adsorption of the *donor* molecules to the substrate. This is in agreement with previous work on small water clusters adsorbed on metallic substrates, such trimers on Pt[47] and hexamers on Cu.[13]

In summary, from the calculations presented in this section we can conclude that, quite independently of the structural details of the surface, the interaction with a Ru substrate can give rise to a significant stabilization of the H-bonds between water molecules when the bonding to the substrate takes place through the donor molecule. Presumably this is also the case for other metallic substrates, as the calculations in References [13, 44], respectively for Pt and Cu, indicate.

**Conclusions**

Motivated by the large stability of a water dimer on O(2x2)/Ru(0001) as recently calculated by some of us,[33] we have studied the role of substrate induced cooperative effects in the adsorption of water. We have performed DFT calculations for two sets of systems: a water dimer on O(2x2)/Ru(0001) and a water dimer interacting with a small Ru cluster. A significant strengthening of the intermolecular H-bond, accompanied by the corresponding weakening of the donating O-H covalent bond, is observed for the studied systems. The reinforcement of the H-bond is due to the strong polarization induced by the interaction with the metallic substrate in the donor molecule. We thus confirm the importance of cooperative effects on the water adsorption on metallic substrates, in agreement with the recent findings[11, 13, 44] for water adsorption on Ag, Cu and Pt substrates.

Two important consequences can be extracted from our results:

(i) It is particularly interesting to note that the H-bond reinforcement only takes place when the water donor molecules are the ones directly bonded to the substrate. This provides a simple route to propose sensible structural candidates for water bilayers on metallic substrates. Therefore, in general, those structures in which the binding to the metal takes place through the donor molecule can be expected to be significantly more stable that those in which the acceptor water molecules are attached to the substrate. For example, this provides a very simple rationalization of the results found in Ref.[45].

(ii) Our results for the O(2x2)/Ru(0001) surface show that, even in the presence of co-adsorbates like Oxygen atoms, the metal-water interaction is the main source of substrate-induced polarization of the water molecules, and thus of substrate-induced cooperative effects. This is quite interesting since the Oxygen atoms in the surface can cause an important polarization of the substrate and are able to form H-bonds with the adsorbed water molecules. Still the water-metal interaction seems to have a key role in the appearance of cooperative effects.

In the present work, we have analyzed in detail the role of substrate-induced cooperative effects on the adsorption of water on clean and decorated metallic substrates. Such cooperative effects will be present for all cases in which the water-substrate interaction can cause a significant polarization of the molecule. Surfaces of ionic crystals represent good substrate examples where such effects could be expected. Indeed, the high stability of the structures based on the dimer-motif on NaCl(001), as compared to those based on well adsorbed monomers, can be interpreted as a manifestation of cooperative effects coming into play.[15, 34] However, in those cases the energy difference between both configurations was much smaller than that found in the present work for adsorption on Ru surfaces.

We can now speculate with the implications of our findings for the chemistry of water. Since the reinforcement of the H-bond and polarization of the molecule typically takes place at the expense of weakening the donating covalent O-H bond, we can expect a decrease of the energy barrier for partial dissociation for some of the structures studied in this paper and, in general, for the water bilayers and other buckled H-bonded networks on metallic substrates. In particular, the behaviour found for the O(2x2)/Ru(0001) seems to agree with the experimental observation that co-adsorption of small amount of oxygen with water on Ru(0001) enhance the dissociation of water.[2, 6] Besides the effect of the adsorbed oxygen on the energetics of the dissociation products studied in Ref.[12], one could argue that the presence of oxygen is likely to favour the presence of buckled structures similar to the water dimer studied here and reminiscent of the bilayer. These structures would favour the appearance of important cooperative effects and the reduction of the dissociation barriers. However, further work is necessary to calibrate the actual importance of substrate-induced cooperative effects on the partial dissociation of water on metallic substrates.

**Acknowledgements**
We acknowledge support from Basque Departamento de Educación, UPV/EHU (Grant No. IT-366-07), the Spanish Ministerio Innovación y Ciencia (Grant No. FIS2007-66711-C02-00) and the ETORTEK research program funded by the Basque Departamento de Industria and the Diputación Foral de Guipúzcoa. MVF-S acknowledges support from DOE grant DE-FG02-09ER16052.